# Observations of Near-Earth Optical Transients with the Lomonosov Space Observatory

V. M. Lipunov[1, 2] *, E. S. Gorbovskoy[1], V. G. Kornilov[1, 2], V. V. Chazov[1],
M. I. Panasyuk[3], S. I. Svertilov[3, 2], I. V. Yashin[3], V. L. Petrov[3],
V. V. Kallegaev[3], A. A. Amelushkin[3], and D. M. Vlasenko[1, 2]

1    Sternberg Astronomical Institute, Lomonosov Moscow State University,
Moscow, 119991 Russia
2 Lomonosov Moscow State University, Physics Department,
Moscow, 119234 Russia
3    Skobel'tsyn Institute of Nuclear Physics, Lomonosov Moscow State University,
GSP-1, Moscow, 119234 Russia
Received November 2, 2017; in final form, January 26, 2018

**Abstract**—The results of observations with the MASTER-SHOK robotic wide-field optical cameras onboard the Lomonosov Space Observatory carried out in 2016 are presented. In all, the automated transient detection system transmitted 22 181 images of moving objects with signal-to-noise ratios greater than 5 to the Earth. Approximately 84% of these images are identified with well-known artificial Earth satellites (including repeated images of the same satellite) and fragments of such satellites (space debris), according to databases of known satellites. The remaining 16% of the images are relate to uncatalogued objects. This first experience in optical space-based monitoring of near-Earth space demonstrates the high efficiency and great potential of using large-aperture cameras in space, based on the software and technology of the MASTER robotic optical complexes (the Mobile Astronomical System of TElescope-Robots (MASTER) global network of robotic telescopes of Lomonosov Moscow State University).

## 1. INTRODUCTION

The Russian multipurpose Lomonosov Observatory was launched on April 28, 2016 from the Vostochnii spaceport [1]. This is a multichannel scientific complex for studies of various phenomena in the Universe over essentially the entire electromagnetic spectrum and at the energies of cosmic-ray particles. This is the first Russian satellite using the MASTER-SHOK (which we will refer to as SHOK) onboard robotic wide-field orbital cameras for space research [2].

The MASTER-SHOK cameras mounted on board the Lomonosov Space Observatory are in-tended for two objectives. First, the very wide field of view of these cameras (2 × 1000 sq. degree) en-ables optical measurements in support of the BDRG gamma-ray detectors of the X-ray and gamma-ray detector unit. The SHOK cameras, which intercept about 10% of the gamma-ray detections, are able to record the optical emission before, during, and close to gamma-ray bursts (GRBs), which is virtually impossible in alert-mode observations [3]. Although the SHOK cameras were not able to perform such observations during the first year of the Lomonosov mission (2016), similar cameras mounted on the robotic telescopes of the MASTER global network [4] carried out unique optical observations simultaneously with GRB 160625A [3].

Secondly, the possibility of monitoring near-Earth space directly from space is of special interest, since full processing of the MASTER-SHOK images distinguishing stationary and moving catalogued and unknown optical sources can be performed in real time (1 minute after each CCD readout) directly on-board the Lomonosov Space Observatory. Such observations have now been realized for the first time in the framework of the Russian Space Program, as will be reported in this paper. While the image scale of 36 does not enable us to determine coordinates with astrometric accuracy, we emphasize that the primary goal of this experiment was the detection of unknown or previously unmonitored objects in near-Earth orbit. It is possible to determine exact coordi-nates, and therefore exact trajectories, with the imple-mented scale of the MASTER-SHOK cameras since information about the detection will be operatively transmitted to ground facilities capable of carrying out accurate astronomical observations.

* E-mail: lipunov2007@gmail.com

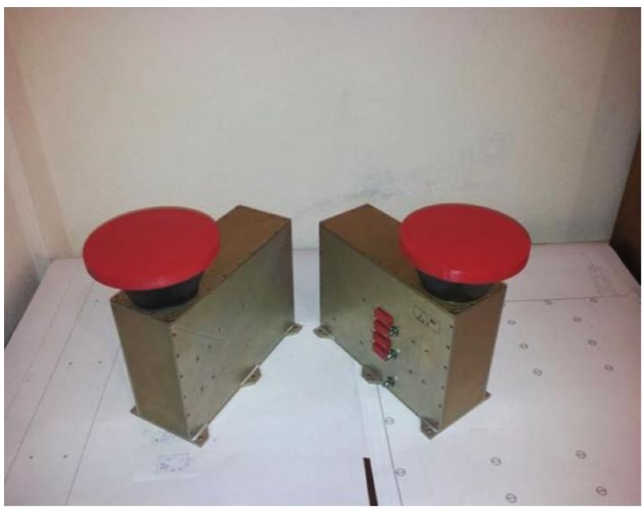

**Fig. 1.** The MASTER-SHOK 1 and MASTER-SHOK 2 devices prepared for installation on the Lomonosov spacecraft.

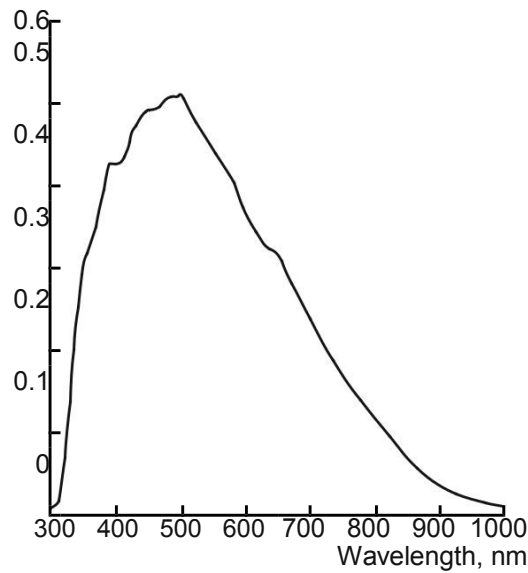

**Fig. 2.** Quantum efficiency curve for the Kodak Truesense KAI-11002 CCD.

## 2. EQUIPMENT

The MASTER-SHOK device (Fig. 1) installed onboard the Lomonosov Space Observatory [1] is a fast, ultra-wide field camera, which enables the recording of optical emission with the maximum magnitude in a single frame with an exposure time of 0.2 seconds.

The device is intended for sky monitoring with a wide field of view and the detection and localization of fast, time-variable (transient) sources on the celestial sphere at optical wavelengths, including synchronous recording of the optical emission from regions on the celestial sphere containing sources of GRBs recorded by the BDRG device, implemented using a control signal (trigger) from the BDRG and/or the UBAT device [1].

The BDRG [5] and UBAT (UFFO Burst Alert Telescope) [6] devices are based on different principles for the detection of hard X-ray emission onboard the Lomonosov spacecraft. Each is able to record GRBs and transmit the corresponding signals to the MASTER-SHOK devices in real time. The BDRG [5] consists of three identical NaI(Tl)/CsI(Tl) scintillator detector units directed toward different sectors of the sky, which are able to determine the position of a new source on the sky from the differences between the count rates of each of the detectors. The UBAT [6] is part of the Ultra-Fast Flash Observatory (UFFO) onboard the Lomonosov; this is an X-ray telescope with a coded mask.

The Lomonosov Space Observatory is equipped with two identical devices, SHOK 1 and SHOK 2. Each of these devices represents a monoblock consisting of an optical recording unit, electronics assembly, mechanical design elements, and the housing.

The electronics assembly consists of a recording camera with an optical lens attached to it. The linear dimensions of the camera are 110 × 66 × 66 mm, and the dimensions of the lens are 69 × 48 mm. The camera is equipped with a Kodak KAI-11002 TrueSense CCD array with 4008 × 2672 pixels and a pixel size of 9 $\mu$m. The camera records optical emission with a maximum quantum efficiency of 50% at 500 nm wavelength. The efficiency curve of the camera is shown in Fig. 2. The CCD camera enables recording and image read-out at a rate of five frames per second at full resolution. The read-out is performed without time gaps between the images.

A modified 50 mm f/1.2 AI-S Nikon lens specially adapted to space-flight conditions is attached to the camera using a special flange. The focal length of the lens is 50 mm, and its aperture ratio is $f/d = 1.2$. These parameters together with those of the camera provide a field of view of 1000 square degrees. At the same time, the seeing of the images over the field of view is FWHM (full width at half maximum) ~ three pixels.

The optical recording unit has a sensitivity that enables the detection of optical radiation corresponding to magnitudes of 9–10$^m$ in a single frame, in the mode with the maximum frame rate (five frames per second). When 100 frames are summed, the sensitivity reaches 12–13$^m$, taking into account the image seeing, the quantum efficiency of the camera, the noise level, etc.

The camera was adapted to space-flight condi-tions based on the results of complex thermal vac-uum, vibration, and radiation testing. To prevent overheating of the CCD image processors in vacuum, the camera is filled with a PK-68-type heat sink compound, TU 38-103508-81, with a boron nitride filler. The camera contains temperature sensors. The diaphragm is separate from the lens, and the lens focus rings have been improved to enable secure fix-ation of the focus. Ground checking images of the sky provide an average seeing (FWHM) of 2.5 pixels, with a maximum size of the image edge of 3.6 pixels. Such images are nearly optimal for detection and acceptable astrometrical coordinate measurement.

The information recorded by the camera enters the electronics assembly through a standard IEEE 802.3 1000baseT gigabit interface.

The electronics assemblies of each of the SHOK 1 and SHOK 2 devices include the following modules:

• A Lippert Cool RoadRunner-945GSE processor module with an Intel Atom 1.6 GHz processor. This performs the data acquisition, processing, analysis, storage, and retrieval, and controls the SHOK device operation.

• A RTD CM17222ER module with two inde-pendent ethernet interface channels provides ethernet communication with the primary and stand-by BI unit half-sets at a rate of up to 1 GBit/s.

• A CAN RTD ECAN527 interface module that communicates with the primary and stand-by BI unit half-sets through the CAN protocol.

• A RTD IPWR104HR-60 secondary power source that converts the input voltage (18 V–36 V) to the voltage required for the operation of the device modules.

In general, the electronics assembly allows con-tinuous surveying of the sky, and provides for frame rotation and storage, minimum frame processing, and preparation of the data for transmission to the Earth (see Section 3 for details).

The appearance of the SHOK devices is shown in Fig. 1. Each SHOK device is powered by an external circuit with a voltage of 24 V. The main characteris-tics of the SHOK devices are given in the Table 1.

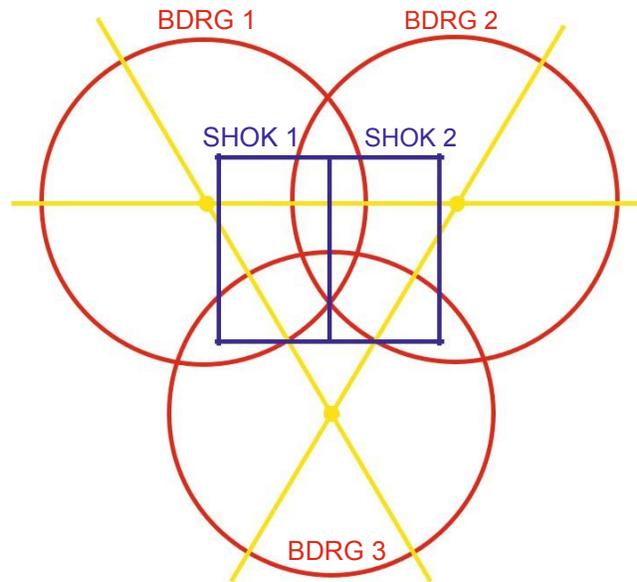

**Fig. 3.** Intersection of the field of view of the instruments on board the Lomonosov mission. The red circles (60° in diameter) indicate the maximum sensitivity zone of each of the three BDRG devices, and the blue rectangles (~26° × 40°) are the fields of view of the MASTER-SHOK 1 and 2 devices. The yellow lines connect the centers of the BDRG detectors.

## 3. MODES OF OPERATION OF THE SHOK DEVICES

The SHOK field of view is in the GRB detec-tion area of the BDRG device located onboard the Lomonosov spececraft (Fig. 3). In fact, the cameras are continuously acquiring "film", part of which may be transmitted back to the Earth upon detection of a GRB. The cameras are securely fixed to the space-craft platform and rotate with it. Thus, the whole sky is scanned. Image processing is carried out for optical-transient searches between GRBs, with real-time differential frame processing performed onboard. As a result, the difference frame displays all moving objects (nearby asteroids, satellites, space debris), objects that are strongly brightening, and uncata-logued objects. Information about all new discovered objects and images of the brightest of these are sent to the Earth for further analysis during the day. The Lomonosov is the first Russian satellite with an on-board internet network. The mission payload includes several GlobalStar modems, providing narrow, but crucially constant, contact with the Earth. Fur-thermore, this enables real-time alerts issued for the brightest and most interesting events requiring rapid response from ground facilities, including detections of the nearest asteroids with the SHOK and GRBs with the other detectors sent to ground-based obser-vatories for more detailed investigation. The internet also allows for the consistent availability of current

**Table 1.** SHOK device specifications

| Parameter | Value |
|---|---|
| Weight | 5.5 ± 0.3 kg |
| Dimensions | 137 × 295 × 306 mm |
| Power consumption | 26 W |
| Wavelengths detected | 330–820 nm at the threshold quantum efficiency of 10% |
| Lens diameter | 52 mm |
| Lens focal lenght | 50 mm |
| Aperture ratio | f/1.2 |
| Camera resolution | 4008 × 2672 px = 10.7 Mpx |
| Seeing (FWHM) | 2.5–3 pixels |
| FOV | ~26.5 × 39.9 deg ~1000 sq. degree |
| Frame rate | from 5 frames/second |
| Noise | 30 electrons |
| Depth of the potential well | 60 000 electrons |
| Quantum efficiency | 50% at 500 nm (Fig. 2) |
| Limiting magnitude (single) | $9^{m}$ (0.2 s exposure) |
| Limiting magnitude (co-add) | $12$–$13^{m}$ (20 s exposure) |
| Bit ADC | 12-bit |
| Possible binning | 1, 2, 4, 8 |
| Data flow in normal operational mode (uncompressed) | ~2.3 TBt/day |
| Data flow in normal operational mode (compressed) | ~850 GBt/day |
| Data flow for one GRB (1 min. with compression) | ~500 MBt |
| Data flow for one transient | ~1–5 MBt |

asteroid and satellite catalogs onboard, enabling confirmation of onboard analyses.

MASTER-SHOK is the first experiment with ultra-wide-angle cameras in Earth orbit. Since the development of methods for the detection of hazardous asteroids and space debris from space is of particular interest, the presence of internet onboard the spacecraft will enable updating of the software for data analysis, taking into account experience gained in the course of the mission.

## 4. OPTICAL-TRANSIENT SEARCH MODE

In the optical-transient search mode, the camera continuously surveys the sky with short exposures (0.2 s by default). The frames are transmitted to the electronics assembly for processing, compression, and storage. The electronics assembly performs continuous processing in order to detect bright optical transients. This processing is carried out using a difference scheme, i.e., the previous frame is subtracted from the current one, and new objects are detected in the resulting difference image. Upon the detection of a bright object, its information is stored in a special file. Once an hour, for the ~100 brightest events detected within the hour, frame clips centered on the detected source within a few minutes before and after the event are generated from the initial full-frame image set. After formatting, these clips are saved sep-arately for subsequent transmission to the Earth. The small size of the frame clips allows information about a large number of bright transients to be transmitted to the Earth.

The specialzed SkyCam program performs the acquisition, storage, and processing of the frames from a wide-angle optical channel. At the start-up, the program receives information about the data-acquisition parameters (exposure time, binning, subframes) from a configuration file and initiates the continuous data-acquisition mode. The process of saving frames onto the solid-state drive (SSD) can take quite a long time, which is determined by the disk, the bus ca-pacity, and other poorly determined parameters of the system. Therefore, to prevent omissions in the frame reading, the saving of the frames is carried out in an asynchronous regime.

This also provides the opportunity of compressing the frames using the Rice method with the CFIT-SIO FPACK package. This was chosen as the least time-consuming method for compressing the frames in FITS format, and enables reduction of the file size by more than a factor of two (from 5.3 Mb to 1.7 Mb for $2004 \times 1336 \text{ pixel}^2$ frames). Note that, even though the compression requires additional computing power, it reduces the load to the SSD disk. Thus, the overal reduction in the system performance with the compression is small. In addition, the ac-tivation of the frame compression can increase the information onboard storage time by about a factor of three, expanding the time for requesting the frame output received from the external trigger.

Up to 25 frames can be saved simultaneously. If the number of unsaved files exceeds 25, the data-acquisition program makes a two-second pause for a system dump. This is done to prevent disruption of the stability of the system's operation, taking into account possible wear and unforeseen situations on-board, although no such a situation was observed during tests with the highest possible frame rate (five frames per second and $2 \times 2$ binning), frame com-pression, processing, and an additional load due to the astrometric tie of one of the frames.

The same computer as the one used for the data acquisition is used to process the frames from each wide-angle optical channel. Under these conditions, data acquisition and frame saving are priorities, and computing resources are provided accordingly. Starting any intensive processing operations is not possible in this case.

The reduction process foresees the acquisition of difference frames between consecutive images. In this case, we assume (to speed up the calculations) that the background changes only slightly in 0.2 s, so that the mean background on the difference frame will be zero. The standard deviation from the mean is calculated at the time for the difference frame. A recursive search for the bright related objects is then performed. To facilitate the search, we select only the brightest objects (with signals exceeding the background by $10-15 \sigma$). Such objects may often be found in the wings of the brightest stars in the original frame. Therefore, to confirm our object selection, we check the presence of the object found in a new frame, and its absence in the previous frame. Each object found in this manner is written to a stack of objects found. Every hour, clips of up to $128 \times 128$ pixels are formed for the 100 brightest objects, centering on the object for subsequent transmission to the Earth.

The continuous operation is monitoring using the camcheck program, which keeps track of the modi-fication time of the file /dev/shm/checkcam. If this file is not updated for more than 10 seconds (the maximum possible exposure), the command to restart is sent.

The deleter program monitors free space on the drive and the frame rotation. Files are deleted in series of 1-minute duration by deleting the corresponding folder in the archive.

## 5. COLLABORATION BETWEEN SHOK AND THE MASTER GLOBAL ROBOTIC TELESCOPE NETWORK

Current astrophysical and applied investigations require operative communication between spacecraft and ground-based devices—information centers and robotic devices designed for monitoring and refining the coordinates of detected objects. For example, when a wide-field camera in orbit discovers a new object and immediately transmits this information to more powerful tracking units, the discovered object can be rapidly identified and its trajectory fully deter-mined with the required accuracy (see the description of the operation of the MASTER global network, for which real-time automatic communication with the Swift, Fermi, Integral, and other spacecraft [3, 9, 10, 12] is already implemented). Experience with this type of operation shows a drastic improvement of the accuracy with which transients—short-lived or fast moving objects—can be localized. For example, the accuracy of coordinate determination is improved a thousandfold as a result of such real-time inter-action. This enables, for example, the discovery of optical bursts in regions (error boxes) with areas of hundreds of square degrees around Fermi GRBs [12] or gravitational-wave bursts such as LIGO/Virgo GW170817 associated with a kilonova [9, 10], mak-ing it possible to determine the source coordinates with accuracies to fractions of an arcsecond. This removes problems associated with the crude accuracy of short-focus, wide-field optical systems in orbit. It is obvious that this coordinated operation of space and ground-based instruments significantly decreases the cost of the space apparatus (by several orders of mag-nitude).

The MASTER global robotic network includes both a set of alert telescopes that can rapidly respond to alerts received from the Gamma-ray Coordinates Network[1] or directly via socket connection from LIGO/Virgo, IceCube, and ANTARES—enabling the detection of optical emission from sources with various physical natures (GRBs [3], neutrino events [8], gravitational-wave events [9–11])—and a network of robotic telescopes independently surveying the sky to search for optical transients, which can send alerts to other observatories for further spectral observations and more detailed photometry. In this way, the MASTER network is able to detect more then ten different types of optical transients: prompt and afterglow radiation associated with GRBs (in-cluding those with large coordinate uncertainties [12]), supernovae (including those important for studies of cosmological parameters [13]), novae [14], quasar flares, decaying objects ($\varepsilon$ Auriga objects [15]), polars and other types of cataclysmic vari-ables, comets and asteroids (including potentially hazardous objects), and fast transients with unknown natures [16]. The efficiency of the MASTER network has been confirmed by hundreds of observations of GRB error boxes, with dozens of optical sources detected at very early stages (up to 100 seconds after the trigger), as well as the detection of early optical polarization [17] and polarization of the prompt GRB emission [3].

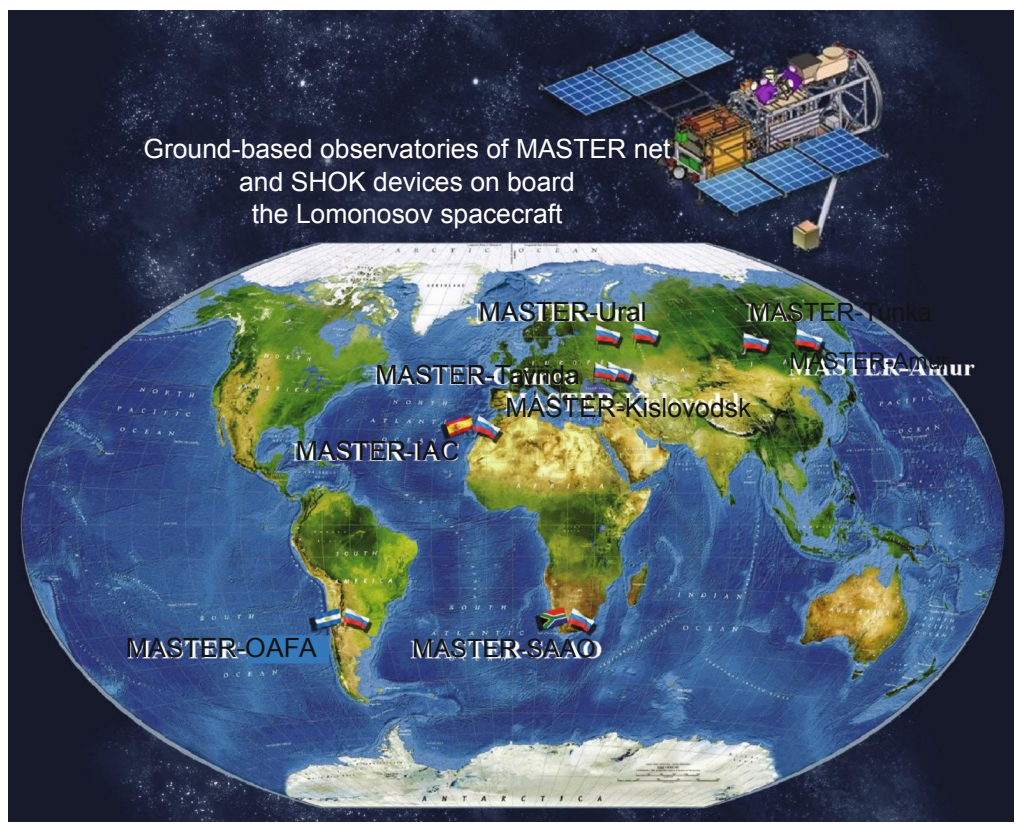

**Fig. 4.** Locations of the ground-based observatories of the MASTER global robotic network and the MASTER-SHOK devices on board the Lomonosov mission (two purple blocks at the top of the spacecraft).

The MASTER global robotic network[2] (see Fig. 4) includes eight observatories: MASTER-Amur, MASTER-Tunka, MASTER-Ural, MAS-TER-Kislovodsk, MASTER-Tavrida (Russian Fed-eration), MASTER-SAAO (South Africa), MAS-TER-IAC (Spain, Canarias), and MASTER-OAFA (Argentina) [4]. Each of these observatories is able to survey 128 $\deg^2$/hour with a limiting magnitude of $20^m$ on dark, moonless nights, and each is equipped with two wide-angle optical telescopes with a total field of view of 8 $\deg^2$, a 4098 × 4098 pixel CCD camera with a scale of 1.85 /pixel, and a photometer with Johnson BVRI filters and polarizers [5, 16].

The operation of the MASTER-VWF very-wide-field cameras began in 2005 at the MASTER-Kislovodsk observatory. In all, 12 cameras similar to the MASTER-SHOK cameras are currently op-erating: two on each of the MASTER observatories except for MASTER-Ural.

---

1 http://gcn.gsfc.nasa.gov/

2 http://observ.pereplet.ru/

All alerts from onboard the Lomonosov Space Observatory are transmitted to the Earth through the Globalstar channels with a minimum time delay, making it possible for any telescope in the world to observe a GRB registered by the Lomonosov mis-sion. However, independent of this, observations with the MASTER network are extremely important for several reasons. First, observations of GRBs from the Lomonosov will be a priority for the MASTER network. Second, it is expected that most GRBs will be registered by the Lomonosov BDRG device, and so will have fairly wide error boxes (one square degree or more). Therefore, the primary task of the optical observations is to detect GRB optical counterparts, which cannot be done using narrow-field and/or non-robotic telescopes. Among existing projects, the Lomonosov BDRG field of view is most similar to that of the FERMI GBM [20]. To date, there are only two teams in the world capable of detecting GRB optical counterparts within the FERMI GBM error boxes (hundreds of square degrees in size)— MASTER [12] and iPTF [21].

Numerous observations of extremely bright (brighter than $10^m$) GRBs are currently available, and each event is unique. For example, GRB 160625B was detected by the ground-based analog of SHOK— the MASTER VWF with a limiting magnitude of $8^m$ [3]; GRB 990123 [22] was the first $8^m$ GRB for which prompt optical emission was detected; the very bright ($6^m$ at its maximum) GRB 080319B [23] is the only GRB with an optical light curve having a resolution exceeding a second; and there is also the equally bright burst GRB 130427 with magnitude $7^m$ [24]. The detection of such bright bursts has been a great success. It is extremely difficult to observe bursts in such a bright phase from the Earth for a number of objective reasons: the day and night cycle, weather fluctuations, and the fact that the field of view of a space gamma-ray telescope (i.e., the coordinates of the detected GRB) can be below the horizon at the observation point.

The SHOK device is a first attempt to use a very-wide-field optical camera in conjunction with gamma-ray detectors on board the same satellite in order to eliminate these factors. We expect that the two continuously operating SHOK cameras will record GRBs at a rate of about one burst per year. We also expect the recording of bright, fast optical transients prompted by the internal trigger from SHOK as a result of the onboard image processing.

The most interesting of these sources are observed and confirmed by the MASTER network. A prime example is the observation and discovery of the prompt optical emission from GRB 161017A, detected by the Lomonosov, using the MASTER network [25].

## 6. OBSERVATIONS OF NEAR-EARTH OBJECTS

Although the onboard processing of the SHOK data is generally intended for searches for point-like, stationary, rapidly varying objects (such as prompt GRBs), the SHOK cameras are also useful for ob-servations of artificial Earth satellites. As was noted above, in the normal operational mode, the program will find all bright, rapidly variable objects. Due to their rapid motion, satellites form short-term flashes in a particular place, which (without additional algo-rithms for the filtration of optical transients) are man-ifest as bright transients. During the flight-testing interval up to the end of 2016, the additional program filters were turned off, and the SHOK detectors de-tected, recorded, and transmitted to the Earth images of artificial Earth satellites and space debris.

As was already noted above (Section 3), to reduce the data transmitted in the optical-transient search mode, clips centered on objects found are transmitted to the Earth instead of the full frames. In the context of observations of artificial Earth satellites, the data transmitted from the SHOK detectors represent a "film" of the detected object's passage through the detection point (see Fig. 5). We analyzed all data transmitted to the Earth from the first switch-on of the SHOK detectors on April 15, 2016 up to Novem-ber 12, 2016. Figure 6 shows a time diagram of the operation of the SHOK 1 and 2 devices indicating the number of satellites detected.

We used only data from the onboard processing to demonstrate the direct, onboard detection capabilities of the spacecraft. The full-frame images obtained in the alert mode were not analyzed. Data for more than 300 000 recorded objects were obtained in the time interval indicated above. All of these were analyzed on the Earth. In all, these included 22 181 traces of satellites detected in automatic searches.

According to telemetry data and taking into ac-count the positioning of the SHOK devices relative to the Lomonosov spacecraft, an object's coordinates can be determined with an accuracy of $0.5°$, even without an additional astrometric tie. However, de-spite the extremely small size of the clips (128 × 128 pixels), for more than 85% of the objects, there are more than three stars with sufficiently well known initial positions (∼30) even in this limited field of view, which is sufficient to provide the astrometric tie and improve the accuracy of the coordinate de-termination to 15–20. After the astrometric tie, we identified each of the detected objects using the NORAD catalog, taking into account the positions of the SHOK detectors on the Lomonosov spacecraft in the Earth's orbit.

Among the 22 181 moving objects (satellites) found, 10 711 objects (∼48%) were reliably identified with known objects from the NORAD database in the automated regime, while 7876 objects (∼35%) had no reliable identification, and the remaining 3594 objects (∼16%) were identified with known objects according to their coordinates, but display differences in the directions and speeds of their motion.

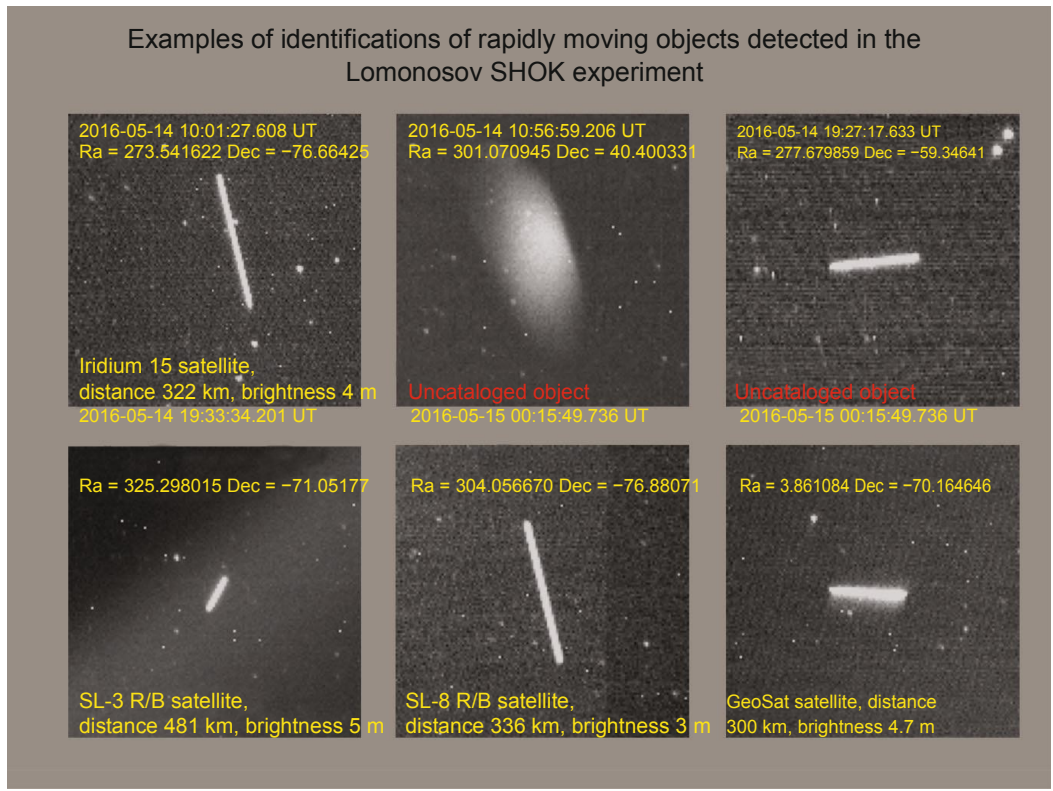

**Fig. 5.** Examples of cataloged and uncataloged satellites detected with the SHOK devices on board the Lomonosov space observatory.

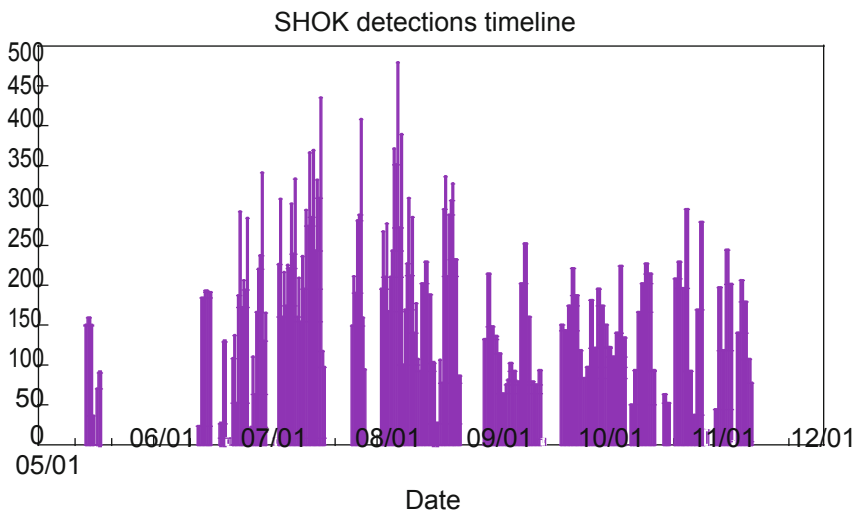

**Fig. 6.** Time diagram of the operation of the SHOK 1 and 2 devices during the flight testing from May 14 to November 12, 2016. The numbers of satellites detected with the SHOK 1 and 2 devices on specific dates are shown.

As was described in Section 3, data for only a small fraction of the brightest recorded objects is transmitted to the Earth. Accordingly, these are the brightest and nearest objects. Figure 7 shows the distribution of the distances of the identified objects from the Lomonosov mission. Satellites can be detected at the shortest distances with the space camera. For example, the well known fragment of space debris PSLV DEB was observed on March 8, 2016 at 21:46:49 at a distance of 21 km, while the TIANHUI 1 spacecraft was observed at a dis-tance 23 km on the same day at 15:41:58.940. The unidentified objects also include some more exotic and apparently closer objects (Fig. 5, center panel in upper row). Note that the distance range up to 200 km is essentially inaccessible for ground-based observations. The camera is mainly sensitive to ob-jects at distances of 100–500 km from the spacecraft, given the methods used.

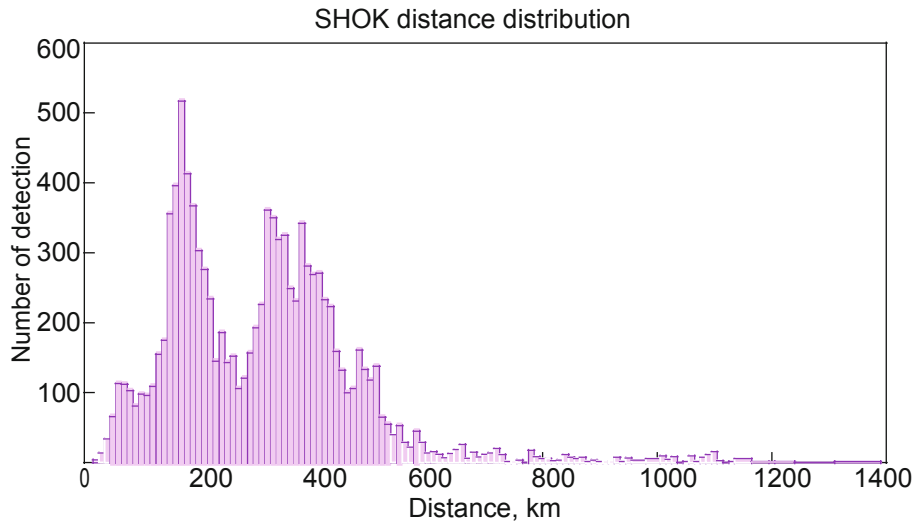

**Fig. 7.** Distribution of the distances of identified objects from the spacecraft. Distances in the range 20–200 km are inaccessible from the Earth.

It follows from the above statistics that even the ultra-wide field-of-view camera onboard the Lomonosov Space Observatory, which is not dedi-cated to studies of moving objects, is able to detect thousands of new low-altitude pieces of space debris.

In practice, only the data for every hundredth of all the objects recorded by the onboard processing are transmitted to the Earth. The onboard process-ing also records much fainter objects, whose number is orders of magnitude greater. If case the orbital complex is equipped with more powerful computers or a significantly improved facility for the storage and transmission of the data obtained to the Earth, this could provide successful monitoring of near-Earth space directly onboard the spacecraft.

## 7. CONCLUSIONS

We have presented the results of the first extensive observations of cosmic objects moving near the Earth (artificial Earth satellites and space debris) from space. Over several months of the flight of the Lomonosov Space Observatory the MASTER-SHOK ultra-wide-field optical cameras obtained about 30 000 thousand images of moving objects. The identifications obtained show the high detection efficiency for previously unknown objects, which are most likely fragments of artificial Earth satellites, or space debris.

Approximately 16% of the observed objects are uncataloged. The experience gained enables us to outline future trends in the development of near-Earth monitoring from space. It is obvious that we should use optical systems with greater diameters, and prob-ably with more cameras, in order to increase the total field of view. This is possible only using groups of low-weight satellites equipped with the same astro-nomical equipment and onboard processing, as well as stable, continuous mobile communication with ground facilities designed for space monitoring.

## ACKNOWLEDGMENTS

This work was supported by the Ministry of Education and Science of the Russian Federation in the framework of the Program of Research and Development for Priority Directions of the Scientific–Technological Complex of Russia for 2014–2020 (project RFMEFI60717X0175).

*Translated by E. Seifina*